\documentstyle[12pt,WSPRL]{article}

\begin{document}

\hfill \vbox{
\halign{&#\hfil\cr
	& NUHEP-TH-95-11 \cr
	& hep-ph/9509210 \cr
	& August 1995 \cr}}
\vspace*{0.9cm}

\centerline{\normalsize\bf NEW INSIGHTS INTO THE PRODUCTION}
\baselineskip=16pt
\centerline{\normalsize\bf OF HEAVY QUARKONIUM\footnote{invited talk
presented at the Symposium on Particle Theory and Phenomenology
at Iowa State University, May 1995.}}

\centerline{\footnotesize ERIC BRAATEN\footnote{address after August 1995:
Department of Physics, Ohio State University, Columbus OH 43210.}}
\baselineskip=13pt
\centerline{\footnotesize\it Department of Physics and Astronomy,
Northwestern University}
\baselineskip=12pt
\centerline{\footnotesize\it Evanston IL 60208 USA}
\centerline{\footnotesize E-mail: braaten@nuhep.phys.nwu.edu}

\vspace*{0.9cm}
\abstracts{
Recent data from the Tevatron have revealed that the production rate of
prompt charmonium is more than an order of magnitude larger than the
best theoretical predictions of a few years ago.
This surprising result can be understood by taking into account
new production mechanisms that include fragmentation and the formation
of charmonium from color-octet $c \bar c$ pairs.
}

\normalsize\baselineskip=15pt
\setcounter{footnote}{0}
\renewcommand{\thefootnote}{\alph{footnote}}

\section{Introduction}

The typical high energy physics conference these days
includes talk after talk showing remarkable agreement
between experiment and theory.  There is an occasional two-sigma
discrepancy, but most such problems will go away if you have the patience
to wait for better data.  However there is one problem where
experimental results have differed from theoretical
predictions by orders of magnitude.
This problem is the production of charmonium at large
transverse momentum at the Tevatron.

\section{Color-singlet Model}

Until recently, the conventional wisdom on the production of heavy
quarkonium was based primarily on the {\it color-singlet model}.\cite{schuler}
In this model, the cross section for producing a charmonium state
is proportional to the perturbative cross section for producing a color-singlet
$c \bar c$ pair with vanishing relative momentum and with appropriate
angular-momentum quantum numbers: $^1S_0$ for $\eta_c$, $^3S_1$ for
$J/\psi$ and $\psi'$, $^3P_J$ for $\chi_{cJ}$, etc.
The color-singlet model has great predictive power.  The cross section for
producing a quarkonium state in any high energy process is predicted in
terms of a single nonperturbative parameter for each orbital-angular-momentum
multiplet.  The nonperturbative factor is $|R(0)|^2$ for S-wave states,
$|R'(0)|^2$ for P-wave states, etc., where $R(r)$ is the radial
wavefunction.  For example, the inclusive differential cross sections for
producing $J/\psi$ and $\chi_{cJ}$ in the color-singlet model have the form
\begin{eqnarray}
d \sigma (\psi + X) &=&
d \widehat{\sigma}(c \bar c(\underline{1},{}^3S_1) + X) \; |R_\psi(0)|^2 ,
\\
d \sigma (\chi_{cJ} + X) &=&
d \widehat{\sigma}(c \bar c(\underline{1},{}^3P_J) + X) \; |R_{\chi_c}'(0)|^2 .
\label{csm-P}
\end{eqnarray}

As the name suggests, the color-singlet model is not a
complete theory of quarkonium production derived from QCD.
The model ignores relativistic corrections, which take
into account the nonzero relative velocity $v$ of the quark
and antiquark.  These corrections may be numerically significant,
since the average value of $v^2$ is only about 1/3 for charmonium and
1/10 for bottomonium.  The color-singlet model
also assumes that a $c \bar c$ pair produced in a color-octet
state will never bind to form charmonium.  This assumption must break down
at some level, since a color-octet $c \bar c$ pair can make a
nonperturbative transition to a color-singlet state by radiating a soft
gluon. The clearest evidence that the color-singlet model is incomplete
comes from radiative corrections.
In the case of S-waves, these can be calculated consistently within the
color-singlet model.  However, in the case of P-waves, the radiative
corrections contain infrared divergences that cannot be factored
into $|R'(0)|^2$.  This problem was first noted in 1976 in connection
with the decays of $\chi_c$ states,\cite{barbieri} but it was solved
only recently.\cite{bbly}  The divergence arises from
the radiation of a soft gluon from either the quark or the antiquark
that form the color-singlet $^3P_J$ bound state.
The infrared divergence can be factored into a matrix element
$\langle {\cal O}^{\chi_c}_8(^3S_1) \rangle$ that is proportional to the
probability for a pointlike $c \bar c$ pair in a color-octet $^3S_1$
state to form $\chi_c$ plus anything.
Thus, perturbative consistency demands that the formula (\ref{csm-P})
of the color-singlet model be modified to take into account
the nonzero probability for a $c \bar c$ pair produced in a
color-octet state to bind to form charmonium:
\begin{eqnarray}
d \sigma (\chi_{cJ} + X) &=&
d \widehat{\sigma}(c \bar c(\underline{1},{}^3P_J) + X) \; |R'_{\chi_c}(0)|^2
\nonumber \\
&& \;+\; (2J+1) \;
d \widehat{\sigma}(c \bar c(\underline{8},{}^3S_1) + X) \;
	\langle {\cal O}^{\chi_c}_8(^3S_1) \rangle .
\label{Pwave}
\end{eqnarray}

The color-singlet model can be used to predict the production rate
of charmonium at large transverse momentum in hadron colliders.
The first thorough treatment of this problem was given by
Baier and R\"uckl in 1981, and their analysis remained the
conventional wisdom for the next decade.\cite{baier-ruckl}
A $\psi$ with large $p_T$ can be produced either directly,
or from a $\chi_{cJ}$ with large $p_T$
that decays via $\chi_{cJ} \to \psi \gamma$,
or by the decay of a $B$ hadron with large $p_T$.
Baier and R\"uckl assumed that the
direct production of charmonium is dominated by the parton processes
that are lowest order in the QCD coupling constant $\alpha_s$.
The relevant parton processes that produce $c \bar c$ pairs at large $p_T$
are $g g \to c \bar c + g$, $g q \to c \bar c + q$,
$g \bar q \to c \bar c + \bar q$, and $q \bar q \to c \bar c + g$.
The cross sections $d \widehat{\sigma}$ for these processes are all of order
$\alpha_s^3$, but they have different dependences on $p_T$.  The only parton
process that produces direct $\psi$ is $g g \to c \bar c + g$, and it
gives a cross section that has the behavior
$d \widehat{\sigma}/d p_T^2 \sim 1/p_T^8$ at large $p_T$.
The dominant parton process for
direct $\chi_{cJ}$ is $g g \to c \bar c + g$,
and it gives $d \widehat{\sigma}/d p_T^2 \sim 1/p_T^6$.
Both of these cross sections fall more rapidly with $p_T$
than typical jet production cross sections, which behave like
$d \widehat{\sigma}/d p_T^2 \sim 1/p_T^4$.
The cross section for $b$ quark production has this scaling behavior when
$p_T \gg m_b$.  Thus the conventional wisdom was that $\psi$'s
at large $p_T$ should come predominantly from $b$
quarks, with direct $\chi_{cJ}$'s being the next most important
source, and direct $\psi$'s being negligible.
This conventional wisdom has been completely overthrown by
recent experimental data from the Tevatron.

\section{Prompt Charmonium at the Tevatron}

In the 1988-89 run of the Tevatron, the cross section for $\psi$
production at large $p_T$ was measured by the CDF collaboration.
They also measured the cross section for $\chi_c$ at large $p_T$.
Assuming the conventional wisdom that $\psi$'s at large $p_T$ come
predominantly from $b$ quarks and from direct $\chi_c$'s, they inferred
that the fraction $f_b$ of $\psi$'s that come from $b$ quarks
was about 60\%.\cite{CDF}
This number was then used to determine the $b$-quark cross
section.  Unfortunately, the conventional wisdom proved to be wrong.
The fraction $f_b$ is actually closer to $15 \%$, and the $b$ quark
cross section is significantly smaller than the result obtained in the
1988-89 run.

The breakthrough came in the 1992-93 run of the Tevatron, with the
installation of a silicon vertex detector at CDF.  This device can be used to
measure the separation between the collision point of the $p$ and $\bar p$
and the point where the $\psi$ decays into leptons.
If a $\psi$ with large $p_T$ is produced by QCD mechanisms, then the leptons
from its decay will trace back to the $p \bar p$ collision point
and the $\psi$ is called {\it prompt}.  Similarly, if a $\chi_c$ is produced by
QCD mechanisms and decays radiatively, the resulting
$\psi$ is also prompt.  On the other hand, the $\psi$'s coming from
$b$ quarks are not prompt.  A $B$ hadron with large $p_T$ will travel a
distance on the order of a millimeter before it decays weakly.  Thus the
leptons from the decay of the $\psi$ will trace back to a vertex with a
measureable displacement from the collision point.

By using the vertex detector to
remove the background from $b$ quarks, CDF was able to study
the QCD production mechanisms directly.  They discovered that
the cross section for prompt $\psi$ production at large $p_T$
is about 30 times larger than predicted.\cite{CDF}
The prompt cross section for $\chi_{c1} + \chi_{c2}$
is also larger than predicted by a
similar factor. The cross section for prompt $\psi'$ at large $p_T$
is three orders of magnitude larger than the theoretical prediction.
The conventional wisdom on prompt charmonium production fails completely
when confronted with this data.  The data can only be explained by new
production mechanisms.

\section{Fragmentation}

Braaten and Yuan pointed out in 1993 that the
dominant production mechanism for charmonium at
sufficiently large $p_T$ must be {\it fragmentation},
the formation of charmonium within the jet initiated by a parton with
large transverse momentum.\cite{braaten-yuan}
This production mechanism had not been
included in any previous theoretical predictions for $p \bar p$ collisions.
The importance of fragmentation should have been realized long ago.
There are factorization theorems of perturbative QCD dating back to
1980 that guarantee that the inclusive production of a hadron
at large $p_T$ is dominated by fragmentation.\cite{sterman}
According to these theorems, the asymptotic form
of the inclusive differential cross section for producing a hadron
$H$ with momentum $P$ is
\begin{equation}
d \sigma ( H(P) + X) \;=\;
\sum_i \int_0^1 dz \; d \widehat{\sigma}(i(P/z)+X) \; D_{i \to H}(z) \;,
\label{frag}
\end{equation}
where $d \widehat{\sigma}$ is the differential cross section for
producing a parton of type $i$ with momentum $P/z$ and $D_{i\to H}(z)$
is a fragmentation function.  It gives the probability that a jet
initiated by parton $i$ includes a hadron $H$ carrying
a fraction $z$ of the parton momentum.  The parton cross sections
$d \widehat{\sigma}$ can be calculated using perturbation theory (up to
parton distributions for the colliding particles if they are hadrons).
All the nonperturbative dynamics involved in the formation of the hadron $H$
is contained in the fragmentation functions.

The factorization theorem (\ref{frag})
applies equally well to heavy quarkonium as to light hadrons.
It is easy to see that the conventional calculations of $\psi$ production
at large $p_T$ did not include fragmentation contributions.
At large $p_T$, the parton cross sections $d \widehat{\sigma}$
in (\ref{frag}) have the scaling behavior
$d \widehat{\sigma}/d p_T^2 \sim 1/p_T^4$ up to logarithmic corrections.
However the leading-order cross sections for $g g \to \psi + g$
and $g g \to \chi_{cJ} + g$ in the color-singlet model behave
asymptotically like $1/p_T^8$ and $1/p_T^6$, respectively.
So where are the fragmentation contributions?
The answer is that they do appear in the color-singlet model,
but only at higher order in perturbation theory.
They appear at next-to-leading order for
$\chi_c$ and at next-to-next-to-leading order for $\psi$.
These higher-order corrections would be prohibitively
difficult to calculate in their entirety, but fortunately
the fragmentation functions can be calculated relatively easily.
The fragmentation functions that are needed for prompt $\psi$
production in $p \bar p$ collisions, including $g \to \psi$, $c \to \psi$,
$g \to \chi_{cJ}$, $c \to \chi_{cJ}$, and $\gamma \to \psi$,
have all been calculated to leading order in $\alpha_s$.\cite{fragfun}
The fragmentation functions for $\psi$ were calculated
in the color-singlet model.  The fragmentation functions for $\chi_{cJ}$
were calculated using the factorization formula (\ref{Pwave}),
which includes a color-octet contribution.

To calculate the fragmentation
contribution to the cross section for prompt
$\psi$ production at large $p_T$ in $p \bar p$ collisions,
the fragmentation functions $D_{g \to H}(z)$ and $D_{c \to H}(z)$
must be folded with parton scattering cross sections for $i j \to g + k$
and for $i j \to c + k, \bar c + k$, respectively.
They must then be convoluted with parton
distributions $f_{i/p}(x_1)$ and $f_{j / \bar p}(x_2)$
for the colliding $p$ and $\bar p$.  The largest contribution
comes from the color-octet term in the fragmentation function for
$g \to \chi_c$.  The fragmentation contribution
increases the theoretical prediction by an order of magnitude,
bringing it to within a factor of 3 of the CDF data.\cite{bdfm}
Given the many sources of uncertainty in the calculation,
this can be considered reasonable agreement.

\section{Color-octet Mechanism}

While fragmentation may explain the CDF data on prompt $\psi$
production, it is not enough to explain the data on prompt $\psi'$'s.
In this case, the conventional wisdom
gave predictions that were 3 orders of magnitude smaller than the data.
When fragmentation is included, the prediction increases by more than
an order of magnitude at large $p_T$, but it remains more than an
order of magnitude below the data.
The big difference between $\psi$ and $\psi'$
is that the $\psi$ signal is fed by the decay of $\chi_c$'s,
but there are no known charmonium states that feed the $\psi'$ signal.

One possible explanation for this $\psi'$ anomaly at CDF is that
the $\psi'$ signal is fed by the decays of higher charmonium states
that have not yet been discovered.\cite{close}  Among the
candidates are D-wave states, higher P-wave states, and $c \bar c g$ states.
The main difficulty with this solution is to
explain how these states can have such a dramatic effect on $\psi'$
production at the Tevatron and not have shown up in any charmonium
experiments at lower energy.

Another possibility is that the solution to the $\psi'$
anomaly lies in a new production mechanism.  Braaten and Fleming
have proposed that the $\psi'$'s come primarily from a color-octet
term in the gluon fragmentation function for $\psi'$.\cite{braaten-fleming}
The basis for this proposal is a general theory of inclusive
quarkonium production that was recently developed by
Bodwin, Braaten, and Lepage.\cite{BBL}  This approach allows one to
calculate not only perturbative corrections to any order in $\alpha_s$,
but also relativistic corrections to any order in $v^2$.  The inclusive
cross section for producing a quarkonium state $H$ satisfies a
factorization formula of the form
\begin{equation}
d \sigma ( H + X) \;=\;
\sum_n d \widehat{\sigma}( c \bar c(n) + X) \; \langle {\cal O}_n^H \rangle,
\label{fact}
\end{equation}
where $d \widehat{\sigma}$ is the inclusive cross section for
producing a $c \bar c$ pair separated by a distance less than $1/m_c$
and in a color and angular-momentum state labelled by $n$.
Since $d \widehat{\sigma}$ depends
only on short distances, it can be calculated using perturbative QCD.
The nonperturbative factor $\langle {\cal O}_n^H \rangle$
in (\ref{fact}) is proportional to the probability for a pointlike
$c \bar c$ pair in the state $n$ to form the  bound state $H$.
It can be defined rigorously as a matrix element in nonrelativistic QCD.
The relative importance of the various terms in (\ref{fact})
can be estimated using scaling rules that tell how the matrix elements
scale with $m_c$ and $v$.

The factorization formula (\ref{fact}) also applies to fragmentation
functions, which have the general form
\begin{equation}
D_{i \to H}(z) \;=\;
\sum_n d_{i \to n}(z) \; \langle {\cal O}_n^H \rangle .
\label{Dfact}
\end{equation}
For example, the process $g \to c \bar c$ produces a $c \bar c$ pair
in a color-octet $^3S_1$ state
and gives rise to a term in the gluon fragmentation function
of order $\alpha_s$:
\begin{equation}
D_{g \to H}(z) \;=\;
{\pi \alpha_s \over 24 m_c^3} \delta(1-z) \;
	\langle {\cal O}_8^H(^3S_1) \rangle.
\label{Dglue}
\end{equation}
This term was included in the fragmentation calculations for
direct $\chi_c$ production, because it is required for
perturbative consistency.  It was not included in the calculations
for the S-wave states $\psi$ and $\psi'$ on the grounds
that the matrix element $\langle {\cal O}_8(^3S_1) \rangle$
is suppressed by $v^4$ relative to the nonperturbative factor
$|R(0)|^2$ in the leading color-singlet term.  However the term
(\ref{Dglue}) may well be numerically important, because the leading
color-singlet term is suppressed by a short-distance factor
$d_n(z)$ that is of order $\alpha_s^3$.
The CDF data on $\psi'$ production can be explained
by including the term (\ref{Dglue}) in the gluon fragmentation function
for $\psi'$ and adjusting the matrix element
$\langle {\cal O}_8^{\psi'}(^3S_1) \rangle$ to fit the
data.\cite{braaten-fleming}
The term (\ref{Dglue}) must also appear in the gluon fragmentation
function for $\psi$, and it may explain the CDF data on direct $\psi$'s
that do not come from $\chi_c$'s.\cite{mangano,cho}

\section{Recent Developments}

Recently, there have been several significant experimental developments
in charmonium at large $p_T$ at the Tevatron.
The CDF collaboration has measured the ratio of the prompt
cross sections for $\chi_{c1}$ and $\chi_{c2}$,\cite{vaia} providing one more
test of our understanding of charmonium production mechanisms.
They have also measured the cross sections for the bottomonium states
$\Upsilon$, $\Upsilon'$, and $\Upsilon''$.\cite{vaia}
This information is valuable,
because the larger mass of the bottom quark changes the relative
importance of the various production mechanisms.

There have also been important developments in theory. The
next-to-leading-order perturbative corrections to the fragmentation
function (\ref{Dglue}) have been calculated by Ma.\cite{ma}
Eventually all the relevant fragmentation functions should be calculated
to next-to-leading order in $\alpha_s$.
Another important development is that
Cho and Leibovich have calculated the
terms in the parton cross sections for $i j \to c \bar c + k$
that correspond to the matrix element
$\langle {\cal O}_8^H(^3S_1) \rangle$.\cite{cho}
They used these results to
extend the calculations of prompt charmonium production down
to moderate $p_T$ and also to calculate the
$\Upsilon$ production rate at the largest values of $p_T$ measured
at the Tevatron.
Ultimately, one would like to extend the theoretical predictions all the
way down to $p_T=0$.  It may be necessary
to include other color-octet matrix elements in the factorization
formula (\ref{fact}) for the cross section at low $p_T$.
It will also be necessary
to resum the effects of soft gluons in order to get a cross section
that vanishes properly at $p_T=0$.

In conclusion, data from the Tevatron is driving us toward a deeper
understanding of the production of heavy quarkonium.
The order-of-magnitude discrepancies between experiment and theory
can be understood by introducing new production mechanisms that are
based on a theory of inclusive quarkonium production derived from QCD.
I believe that we are on our way to a comprehensive description of
quarkonium production in all high energy processes in terms of the quark mass,
the QCD coupling constant, and a few well-defined phenomenological
matrix elements.

This work was supported in part by the U.S. Department of Energy,
Division of High Energy Physics, under Grant DE-FG02-91-ER40684.

\end{document}